\documentstyle[11pt]{article}
\def\agt{
\mathrel{\raise.3ex\hbox{$>$}\mkern-14mu\lower0.6ex\hbox{$\sim$}}
}
\def\alt{
\mathrel{\raise.3ex\hbox{$<$}\mkern-14mu\lower0.6ex\hbox{$\sim$}}
}
\begin{document}

\bibliographystyle{unsrt}    

\Large
{\centerline{\bf Gravitational Radiation }} 

\vspace{2mm}

\large
{\centerline{\bf A New Window Onto the Universe}}
\small\normalsize

\vspace{7mm}

\centerline {\bf Kip S. Thorne}
\small\normalsize

\vspace{3mm}

\centerline {California Institute of Technology, Pasadena, CA 91125 USA}

\vspace{4mm}

\begin{abstract}
A summary is given of the  current status and plans for gravitational-wave 
searches at all plausible wavelengths, from the size of the observable universe
to a few kilometers.  The anticipated scientific payoff from these searches is
described, including expectations for detailed studies of black holes and 
neutron stars, high-accuracy tests of general relativity, and hopes for the 
discovery of exotic new kinds of objects. 
\end{abstract}

\vspace{1mm}

\section{Introduction}\label{introduction} 

There is an enormous difference between gravitational waves, and the
electromagnetic waves on which our present knowledge of the Universe is
based: 
\begin{itemize}
\item
Electromagnetic waves are oscillations of the electromagnetic field that
propagate through spacetime; gravitational waves are oscillations of the
``fabric'' of spacetime itself.
\item
Astronomical electromagnetic waves are almost always incoherent 
superpositions of
emission from individual electrons, atoms, or molecules.  
Cosmic gravitational waves are produced
by coherent, bulk motions of huge amounts of mass-energy---either material
mass, or the energy of vibrating, nonlinear spacetime curvature.
\item
Since the wavelengths of electromagnetic waves are small compared to their
sources (gas clouds, stellar atmospheres, accretion disks,
...), from the waves we can make
pictures of the sources.  The wavelengths of cosmic gravitational waves
are comparable to or larger than their coherent, bulk-moving sources, so
we cannot make pictures from them. Instead, the gravitational waves
are like sound; they carry, in two independent waveforms, 
a stereophonic, symphony-like description of their sources.
\item
Electromagnetic waves are easily absorbed, scattered, and dispersed by matter.
Gravitational waves travel nearly unscathed through all
forms and amounts of intervening matter \cite{300yrs,leshouches}.
\item
Astronomical electromagnetic waves have frequencies that begin at $f\sim
10^7$ Hz and extend on {\sl upward} by roughly 20 orders of magnitude.
Astronomical gravitational waves should begin at $\sim 10^4$ Hz
(1000-fold lower than the lowest-frequency astronomical electromagnetic waves),
and should
extend on {\sl downward} from there by roughly 20 orders of magnitude.
\end{itemize}

These enormous differences make it likely that: 
\begin{itemize}
\item 
The information brought to us by
gravitational waves will be very different from (almost ``orthogonal to'')
that carried by electromagnetic waves; gravitational waves will show us
details of the bulk motion of dense concentrations of energy, whereas
electromagnetic waves show us the thermodynamic state of optically thin
concentrations of matter.
\item
Most (but not all)
gravitational-wave sources that our instruments 
detect will not be seen electromagnetically, and
conversely, most objects observed electromagnetically will never be seen
gravitationally.
Typical electromagnetic sources are 
stellar atmospheres, accretion disks, and clouds of
interstellar gas---none of which emit significant gravitational waves,
while 
typical gravitational-wave sources are black holes and other 
exotic objects which emit no electromagnetic wave at all, and 
the cores of supernovae
which are hidden from electromagnetic view by dense layers of
surrounding stellar gas.
\item
Gravitational waves may bring us great surprises.  
In the past, when a radically new window has been opened onto the
Universe, the resulting surprises have had a profound, indeed
revolutionary, impact.  For example, the radio universe, as discovered in 
the 1940s, 50s and 60s, turned out to be far more violent than the optical 
universe; radio waves brought us quasars, pulsars, and the cosmic
microwave radiation, and with them our first direct observational
evidence for black holes, neutron stars, and the heat of the big bang
\cite{sullivan}.
It is reasonable to hope that gravitational waves will bring a similar
``revolution''.
\end{itemize}

Gravitational-wave detectors and detection techniques have been under vigorous
development since about 1960, when Joseph Weber \cite{weber}
pioneered the field.  These efforts have led to promising sensitivities in
four frequency bands:  
\begin{itemize}
\item
The Extremely Low Frequency Band (ELF), $10^{-15}$ to $10^{-18}$ Hz, in which
the measured anisotropy of the cosmic microwave background radiation places
strong limits on gravitational wave strengths --- and may, in fact, have
detected waves \cite{krauss_white,davis}.  The only waves expected in this band
are relics of the big bang.
\item
The Very Low Frequency Band (VLF), $10^{-7}$ to $10^{-9}$ Hz, in which Joseph
Taylor and others have achieved remarkable gravity-wave sensitivities by the
timing of millisecond pulsars \cite{kaspi_taylor}.  The only expected strong
sources in this band are processes in the very early universe --- the big bang,
phase transitions of the vacuum states of quantum fields, and vibrating or
colliding defects in the structure of spacetime, such as monopoles,
cosmic strings, domain walls, textures, and combinations thereof 
\cite{zeldovich_strings,vilenkin_strings,martin_vilenkin}. 
\item
The Low-Frequency Band (LF), $10^{-4}$ to 1 Hz, in which will operate the Laser
Interferometer Space Antenna, LISA; see Sec.\ \ref{lisa} below.  This is 
the band of
massive black holes 
($M\sim 1000$ to $10^8 M_\odot$) in the distant universe,
and of other hypothetical massive exotic objects (naked singularities, soliton
stars), as well of as binary stars (ordinary, white dwarf, 
neutron star, and black
hole) in our galaxy.  Early universe processes should also have produced waves
at these frequencies, as in the ELF and VLF bands.
\item
The High-Frequency Band (HF), $1$ to $10^4$Hz, in which operate earth-based
gravitational-wave detectors such as LIGO; see Sec.\ \ref{gbint} below.  This 
is the band of stellar-mass black holes ($M \sim 1$ to $1000M_\odot$ and 
of other conceivable stellar-mass exotic objects 
(naked singularities and boson stars) in the distant universe,
as well as of supernovae, pulsars, and coalescing and colliding neutron stars.
Early universe processes should also have produced waves 
at these frequencies, as in the ELF, VLF, and LF bands.
\end{itemize}

\begin{figure}
\vskip 9.pc
\special{hscale=65 vscale=65 hoffset=12 voffset=-13
psfile=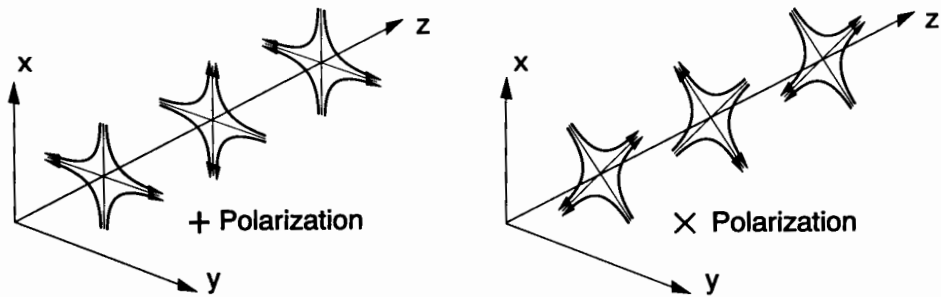}
\caption{The lines of force associated with the two polarizations of
a gravitational wave.  (From Ref. \protect\cite{ligoscience}.)
}
\label{fig:forcelines}
\end{figure}

In this lecture I shall focus primarily on the HF and LF bands, because these
are the ones in which observations are likely to bring us the most new
information.  

\section{Ground-Based Laser Interferometers: LIGO and VIRGO}\label{gbint}

\subsection{Wave Polarizations, Waveforms, and How an Interferometer
Works}
\label{intworks}

According to general relativity theory (which I shall assume to be
correct), a gravitational wave has two 
linear polarizations,
conventionally called $+$ (plus) and $\times$ (cross).  Associated with
each polarization there is a gravitational-wave field, $h_+$ or $h_\times$,
which oscillates in time and propagates with the speed of light.  Each
wave field produces tidal forces (stretching and squeezing forces) on
any object or detector through which it passes.  If the object is small
compared to the waves' wavelength (as is the case for ground-based
interferometers and resonant mass antennas), then relative to the
object's center, the forces have the quadrupolar patterns shown in 
Fig.~\ref{fig:forcelines}.
The names ``plus'' and ``cross'' are derived from the orientations of the
axes that characterize the force patterns \cite{300yrs}. 

\begin{figure}
\center
\vskip 13.2pc
\special{hscale=70 vscale=70 hoffset=15 voffset=-5
psfile=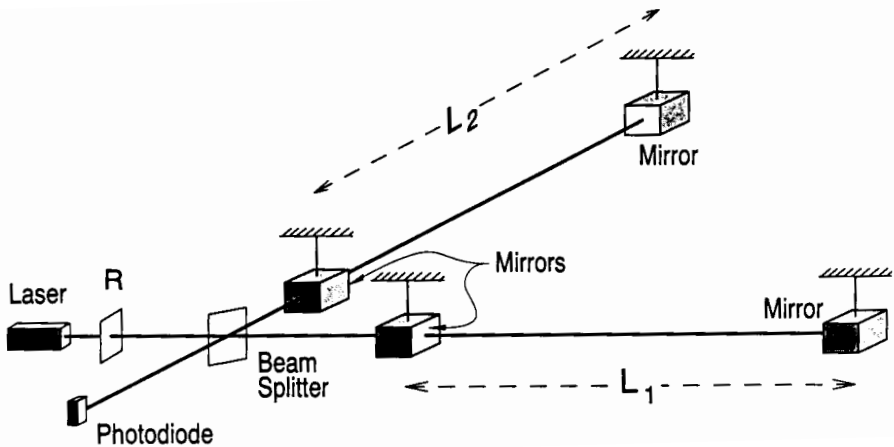}
\caption{Schematic diagram of a laser interferometer
gravitational wave detector.  (From Ref.\ \protect\cite{ligoscience}.)
}
\label{fig:interferometer}
\end{figure}

A laser interferometer gravitational wave detector (``interferometer'' for
short) consists of four masses that hang from vibration-isolated 
supports as shown in Fig.\ \ref{fig:interferometer}, and the indicated 
optical system for 
monitoring the separations between
the masses \cite{300yrs,ligoscience}. Two masses are near 
each other, at the corner of an ``L'', 
and one mass is at the end of each of the L's long arms.  The arm
lengths are nearly equal, $L_1 \simeq L_2 = L$.  When a gravitational
wave, with frequencies high compared to the masses' $\sim 1$ Hz pendulum
frequency, passes through the detector, it pushes the masses back and
forth relative to each other as though they were free from their
suspension wires, thereby changing the arm-length difference,
$\Delta L \equiv L_1-L_2$.  That change is
monitored by laser interferometry in such a way that the variations in
the output of the
photodiode (the interferometer's output) are directly proportional to
$\Delta L(t)$.

If the waves are coming from overhead or
underfoot
and the axes of the $+$ polarization coincide with the arms'
directions, then it is the waves' $+$ polarization that drives the masses, and
$\Delta L(t) / L = h_+(t)$.  More generally, 
the interferometer's output is a linear combination of the two 
wave fields: 
\begin{equation}
{\Delta L(t)\over L} = F_+h_+(t) + F_\times h_\times (t) \equiv h(t)\;.
\label{dll}
\end{equation}
The coefficients $F_+$ and $F_\times$ are of order unity and depend in a
quadrupolar manner on 
the direction to the source and the orientation of the detector \cite{300yrs}.
The combination $h(t)$ of the two $h$'s is called the
{\it gravitational-wave strain} that acts on the detector; and the time
evolutions of $h(t)$, $h_+(t)$, and $h_\times(t)$ are sometimes called
{\it waveforms}.

\subsection{Wave Strengths and Interferometer Arm Lengths}
\label{armlength}

The strengths of the waves from a gravitational-wave source can be 
estimated using the
``Newtonian/quadrupole'' approximation to the Einstein field equations.
This approximation says that $h\simeq (G/c^4)\ddot Q/r$, where
$\ddot Q$ is the second time derivative of the source's quadrupole
moment, $r$ is the distance of the source from Earth (and $G$ and $c$
are Newton's gravitation constant and the speed of light).  
The strongest sources will be highly nonspherical and thus will
have $Q\simeq ML^2$, where $M$ is their mass and $L$ their size, and
correspondingly will have $\ddot Q \simeq 2Mv^2 \simeq 4 E_{\rm
kin}^{\rm ns}$, where $v$ is their internal velocity and
$E_{\rm kin}^{\rm ns}$ is the nonspherical part of their
internal kinetic energy.  This provides us with the estimate
\begin{equation}
h\sim {1\over c^2}{4G(E_{\rm kin}^{\rm ns}/c^2) \over r}\;;
\label{hom}
\end{equation}
i.e., $h$ is about 4 times the gravitational potential produced at Earth by
the mass-equivalent of the source's nonspherical, internal kinetic 
energy---made dimensionless by dividing by $c^2$.  Thus, in order to
radiate strongly, the source must have a very large, nonspherical,
internal kinetic energy.

The best known way to achieve a huge internal kinetic energy is via
gravity; and by energy conservation (or the virial theorem), any
gravitationally-induced kinetic energy must be of order the source's 
gravitational potential energy.  A huge potential energy, in turn, 
requires that the source be very compact, not much larger than its own 
gravitational radius.  Thus, the strongest gravity-wave sources must be
highly compact, dynamical concentrations of large amounts of mass (e.g.,
colliding and coalescing black holes and neutron stars).

Such sources cannot remain highly dynamical for long; their motions will
be stopped by energy loss to gravitational waves and/or the formation of
an all-encompassing black hole.  Thus, the strongest sources should be
transient.  Moreover, they should be very rare --- so rare that to see a
reasonable event rate will require reaching out through a substantial
fraction of the Universe.  Thus, just as the strongest radio waves 
arriving at Earth 
tend to be extragalactic, so also the strongest gravitational waves are 
likely to be extragalactic.  

For highly compact, dynamical objects that radiate in the high-frequency band,
e.g.\ colliding and coalescing neutron stars and stellar-mass black
holes, the internal, nonspherical kinetic energy 
$E_{\rm kin}^{\rm ns}/c^2$ is of order the mass of
the Sun; and, correspondingly, Eq.\ (\ref{hom}) gives $h\sim 10^{-22}$
for such sources at the Hubble 
distance (3000 Mpc, i.e., $10^{10}$ light years); 
$h\sim 10^{-21}$ at 200 Mpc (a best-guess distance for several
neutron-star coalescences per year; see Section \ref{coalescencerates}), 
$h\sim 10^{-20}$ at the
Virgo cluster of galaxies (15 Mpc); and $h\sim 10^{-17}$ in the outer
reaches of our own Milky Way galaxy (20 kpc).  These numbers set the
scale of sensitivities that ground-based interferometers seek to achieve:
$h \sim 10^{-21}$ to $10^{-22}$.  

When one examines the technology of laser interferometry, one sees good
prospects to achieve measurement accuracies $\Delta L \sim 10^{-16}$ cm
(1/1000 the diameter of the nucleus of an atom).  With such an accuracy,
an interferometer must have an arm length $L = \Delta L/h \sim 1$ to 10
km, in order to achieve the desired wave sensitivities, $10^{-21}$ to
$10^{-22}$.  This sets the scale of the interferometers that are now
under construction.

\subsection{LIGO, VIRGO, and the International Gravitational Wave Network}
\label{network}

Interferometers are plagued by non-Gaussian noise, e.g.\ due
to sudden strain releases in the wires that suspend the masses.  This
noise prevents a single interferometer, by itself, from detecting with
confidence short-duration gravitational-wave bursts (though it might
be possible for a single interferometer to search for the periodic
waves from known pulsars).   The non-Gaussian noise can be removed
by cross correlating two, or preferably three or more, interferometers 
that are networked together at widely separated sites. 

The technology and techniques for such interferometers have been under
development for 25 years, and plans for km-scale 
interferometers have been developed over the past 15 years.  An
international network consisting of three km-scale interferometers, at
three widely separated sites, is now in the early stages of 
construction. It includes two sites of 
the American LIGO Project (``Laser Interferometer Gravitational Wave
Observatory'') \cite{ligoscience}, and one site of the French/Italian VIRGO 
Project (named after the Virgo cluster of galaxies) \cite{virgo}.  

\begin{figure}
\vskip 17.7pc
\special{hscale=75 vscale=75 hoffset=10 voffset=-26
psfile=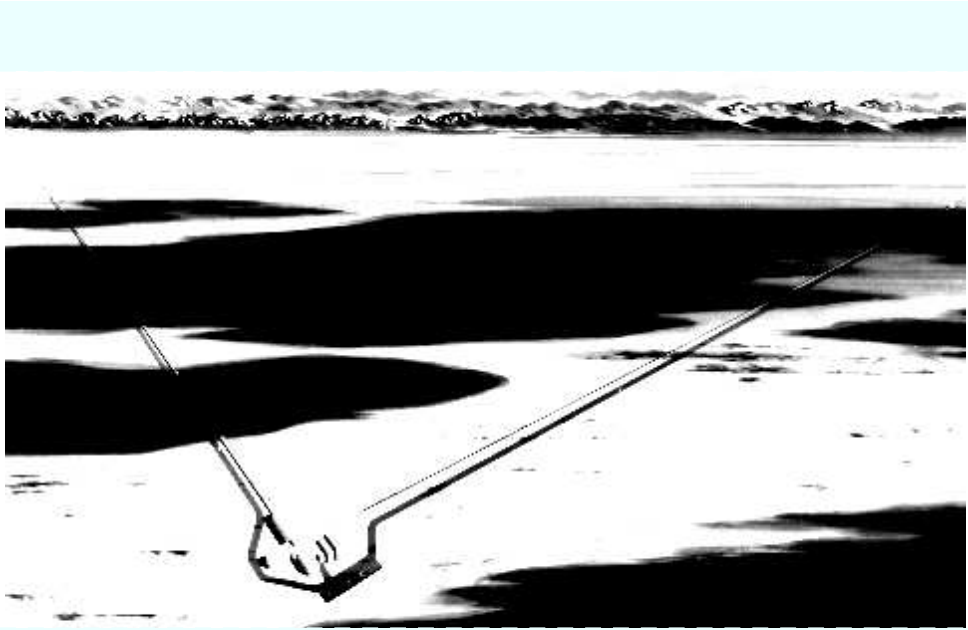}
\caption{Artist's conception of one of the LIGO interferometers.
[Courtesy the LIGO Project.]}
\label{fig:LIGOart}
\end{figure}

LIGO will consist of two 
vacuum facilities with 4-kilometer-long arms, one in Hanford, Washington 
(in the northwestern United States; Fig.\ \ref{fig:LIGOart})
and the other in Livingston, Louisiana (in the southeastern United States).  
These facilities are designed to house
many successive generations of interferometers without the necessity of
any major facilities upgrade; and after a planned
future expansion,
they will be able to house several interferometers at once, each
with a different optical configuration optimized for a different type of
wave (e.g., broad-band burst, or narrow-band periodic wave, or
stochastic wave).  

The LIGO facilities and their first interferometers 
are being constructed by a team of about 80 physicists and engineers at
Caltech and MIT, led by Barry Barish (the PI), Gary Sanders (the
Project Manager), Albert Lazzarini, Rai Weiss, Stan Whitcomb, and
Robbie Vogt (who directed the project during the pre-construction phase).  
Other research groups from many different
universities are contributing to R\&D for {\it enhancements} of the first
interferometers, or are computing theoretical waveforms
for use in data analysis, or are developing data analysis techniques.  
These groups are
linked together by an organization called the {\it Ligo Research Community}.
For further details, see the LIGO World Wide Web Site, 
http://www.ligo.caltech.edu/. 

The VIRGO Project is building one vacuum facility in Pisa, Italy, with 
3-kilometer-long arms.  This facility and its first interferometers are
a collaboration of more than a hundred physicists and engineers
at the INFN (Frascati, Napoli, Perugia, Pisa), LAL (Orsay), LAPP
(Annecy), LOA (Palaiseau), IPN (Lyon), ESPCI (Paris), and the University
of Illinois (Urbana), under the leadership of Alain Brillet and
Adalberto Giazotto.  

\begin{figure}
\center
\vskip23.8pc
\special{hscale=65 vscale=65 hoffset=15 voffset=-5
psfile=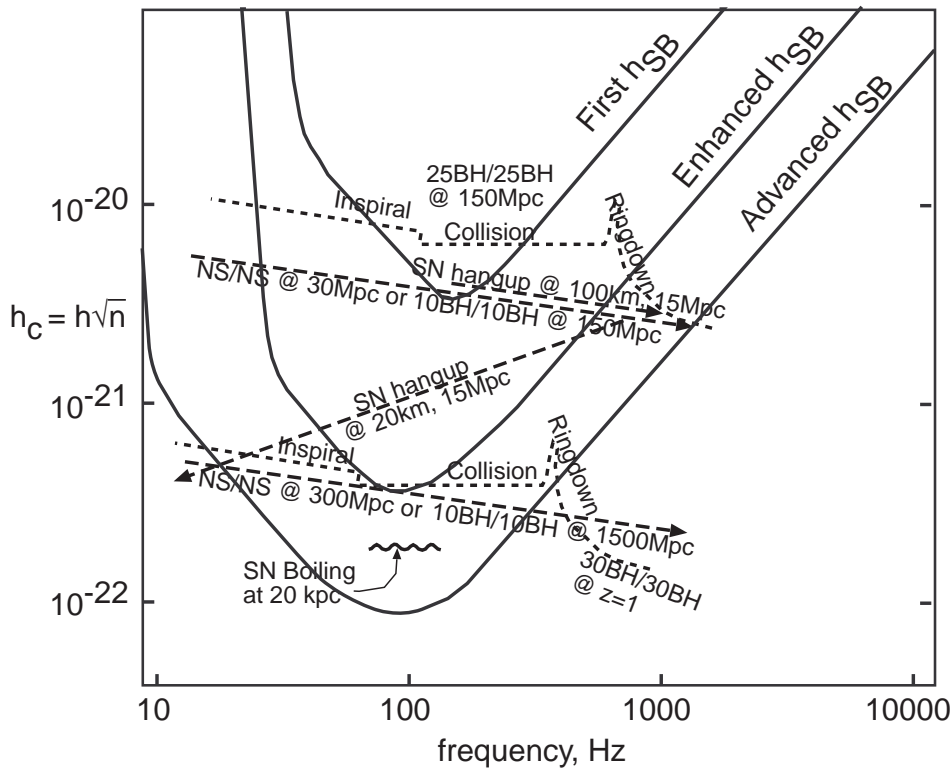}
\caption{LIGO's projected broad-band noise sensitivity
to bursts $h_{\rm SB}$ (Refs.\ 
\protect\cite{ligoscience,RandDproposal}) compared with the 
strengths of the waves from several hypothesized sources.  The
signal to noise ratios are 
$\protect\sqrt 2$ higher than in Ref.\ \protect\cite{ligoscience}
because of a factor 2 error in Eq.~(29) of Ref.\ \protect\cite{300yrs}.  
}
\label{fig:ligosources}
\end{figure}

Both LIGO and VIRGO are scheduled for completion
in the late 1990s, and their first gravitational-wave searches are
likely to be performed in 2001.  Figure \ref{fig:ligosources} shows the 
design sensitivities for LIGO's first interferometers (which are expected to
be achieved in 2001) \cite{ligoscience} and for enhanced versions of those 
interferometers (which is expected to be operating five years or so later)
\cite{RandDproposal}, along with a benchmark sensitivity goal for a more 
advanced interferometer that may operate in the LIGO vacuum system some 
years later \cite{ligoscience,RandDproposal}.  

For each 
interferometer, the quantity shown is the ``sensitivity to bursts'' that come
from a random direction, $h_{\rm SB}(f)$ \cite{ligoscience}.  
This $h_{\rm SB}$ is about 5 times
worse than the rms noise level in a bandwidth $\Delta f \simeq f$
for waves with a random direction and polarization, and about $5\sqrt 5 
\simeq 11$ worse than the the rms noise level $h_{\rm rms}$  for optimally 
directed and polarized waves.   (In much of the literature, the quantity
plotted is $h_{\rm rms} \simeq h_{\rm SB}/11$.)

This $h_{\rm SB}$ is to be compared with the
``characteristic amplitude'' $h_c(f) = h \sqrt{n}$ of the waves from a source;
here $h$ is the waves' amplitude when they have frequency $f$, and $n$ is the
number of cycles the waves spend in a bandwidth $\Delta f \simeq f$ near
frequency $f$ \cite{300yrs,ligoscience}.  Any 
source with $h_c > h_{\rm SB}$ should be detectable with 
high confidence, even if it arrives only once per year.  As examples, Fig.\ 
\ref{fig:ligosources} shows the characteristic amplitudes $h_c$ of
several binary systems made of 1.4$M_\odot$ neutron stars (``NS'') or
10, 25, or 30 $M_\odot$ black holes (``BH''), which spiral together 
and collide under the driving force
of gravitational radiation reaction.  As the bodies spiral inward, their
waves sweep upward in frequency (rightward across the figure along the dashed
lines).  From the figure we see that LIGO's first interferometers should be
able to detect waves from the inspiral of a NS/NS binary out to a distance 
of 30Mpc and from the final collision and merger of a 
$25 M_\odot/25M_\odot$ BH/BH binary out to about 
300Mpc. 

LIGO alone, with its two sites which have parallel arms, will be able to
detect an incoming gravitational wave, measure one of its two waveforms,
and (from the time delay between the two sites) locate its source to within a
$\sim 1^{\rm o}$ wide annulus on the sky.
LIGO and VIRGO together, operating as a {\sl coordinated international
network}, will be able to locate the source 
(via time delays plus the interferometers' beam patterns) 
to within a 2-dimensional error box with size
between several tens of arcminutes and several degrees, depending on
the source direction and on
the amount of high-frequency structure in the waveforms. 
They will also be able to monitor both waveforms $h_+(t)$ and
$h_\times(t)$ (except for frequency components above about 1kHz
and below about 10 Hz, where the interferometers' noise becomes severe). 

A British/German group is constructing a 600-meter interferometer called 
GEO 600 near Hanover Germany \cite{geo}, and Japanese groups, a 300-meter 
interferometer called TAMA near Tokyo.  GEO600 may be a significant player 
in the interferometric network in its early years (by virtue of cleverness and
speed of construction), but because of its short arms it cannot compete
in the long run.  GEO600 and TAMA will both be important development
centers and testbeds for interferometer techniques and technology, and 
in due course they may give rise to kilometer-scale 
interferometers like LIGO and VIRGO, which could significantly enhance the
network's all-sky coverage and ability to extract information from the waves.

\subsection{Narrow-Band Detectors}
\label{narrowband}

At frequencies $f\agt1000$Hz, the interferometers' photon shot noise becomes a
serious obstacle to wave detection.  However, narrow-band detectors specially
optimized for kHz frequencies show considerable promise.  These include a
special ``dual recycled interferometer'' \cite{meers}, 
and huge spherical or icosahedral resonant-mass detectors \cite{tiga}
that are modern variants of Joseph Weber's original ``bar'' detector
\cite{weber}. 

\section{LISA: The Laser Interferometer Space Antenna}
\label{lisa}

The {\it Laser Interferometer Space Antenna} (LISA) is 
the most promising detector for gravitational-waves in the low-frequency band,
$10^{-4}$--$1$ Hz (10,000 times lower than the LIGO/VIRGO high-frequency band).

LISA was originally conceived (under a different name) by Peter Bender of the
University of Colorado, and is being developed by an international team 
led by Karsten Danzmann of
the University of Hanover (German) and James Hough of Glasgow University (UK).
The European Space Agency tentatively plans to fly it sometime in the
2014--2018 time frame as part of ESA's Horizon 2000+ Program of large space
missions.  If NASA contributes significantly to LISA, then the flight could be
sooner.

\begin{figure}
\vskip15.8pc
\special{hscale=70 vscale=70 hoffset=33 voffset=7
psfile=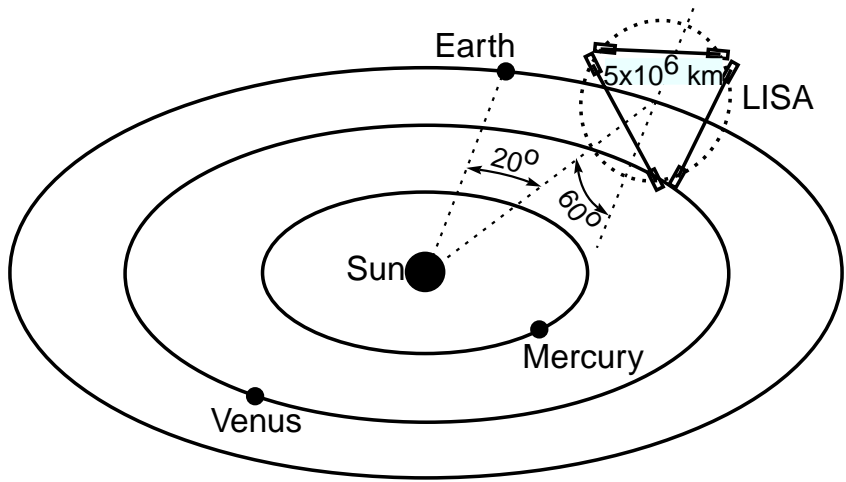}
\caption{LISA's orbital configuration, with LISA magnified in arm length
by a factor $\sim 10$ relative to the solar system.
}
\label{fig:lisa_orbit}
\end{figure}

As presently conceived, LISA will consist of six compact, drag-free
spacecraft (i.e. spacecraft that are shielded from buffeting by solar
wind and radiation pressure, and that thus move very nearly on geodesics of
spacetime).  All six spacecraft would be launched simultaneously by a 
single Ariane rocket. They
would be placed into the same heliocentric orbit as the Earth
occupies, but would follow 20$^{\rm o}$ behind the Earth; cf.\ Fig.\ 
\ref{fig:lisa_orbit}.  The spacecraft would fly in pairs, with each pair
at the vertex of an equilateral triangle that is inclined at an angle of 
60$^{\rm o}$ to the Earth's orbital plane. The triangle's arm length would be 5
million km ($10^6$ times larger than LIGO's arms!).  The six spacecraft would
track each other optically, using one-Watt YAG laser beams.  Because of 
diffraction
losses over the $5\times10^6$km arm length, it is not feasible to
reflect the beams back and forth between mirrors as is done with LIGO.
Instead, each spacecraft will have its own laser; and the lasers will be
phase locked to each other, thereby achieving the same kind of
phase-coherent out-and-back light travel as LIGO achieves with mirrors.
The six-laser, six-spacecraft configuration thereby functions as three,
partially independent but partially redundant, 
gravitational-wave interferometers.

\begin{figure}
\vskip21.0pc
\special{hscale=65 vscale=65 hoffset=20 voffset=-10
psfile=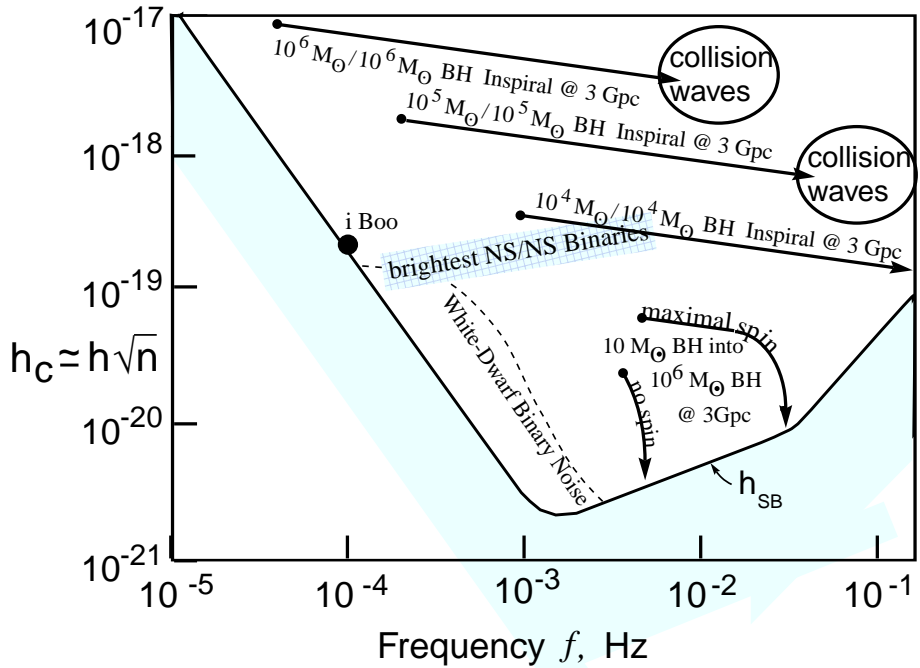}
\caption{LISA's projected sensitivity
to bursts $h_{\rm SB}$, compared with the strengths of the waves from
several low-frequency sources. 
}
\label{fig:lisa_noise}
\end{figure}

Figure \ref{fig:lisa_noise} depicts the expected noise and sensitivity of
LISA in the same language as we have used for LIGO (Fig.\
\ref{fig:ligosources}):  
$h_{\rm SB} =  5\sqrt5 h_{\rm rms}$ is the sensitivity
for high-confidence detection ($S/N=5$) of a broad-band burst coming from 
a random direction, assuming Gaussian noise.   

At frequencies $f\agt 10^{-3}$Hz, LISA's noise is due to photon counting
statistics (shot noise).  The sensitivity curve steepens at $f\sim
3\times10^{-2}$Hz because at larger $f$ than that, the waves' period is
shorter than the round-trip light travel time in one of LISA's arms. 
Below $10^{-3}$Hz, the noise is due to buffeting-induced random motions 
of the spacecraft
that are not being properly removed by the drag-compensation system.
Notice that, in terms of dimensionless amplitude, LISA's sensitivity is
roughly the same as that of LIGO's first interferometers (Fig.\ 
\ref{fig:ligosources}), but at 100,000 times lower frequency.  Since
the waves' energy flux scales as $f^2 h^2$, this corresponds to $10^{10}$
better energy sensitivity than LIGO.

LISA can detect and study, simultaneously, a wide variety of different
sources scattered over all directions on the sky.  The key to
distinguishing the different sources is the different time evolution of
their waveforms.  The key to determining each source's direction, and
confirming that it is real and not just noise, is the manner in which
its waves' amplitude and frequency are modulated by LISA's complicated
orbital motion---a motion in which the interferometer triangle rotates around
its center once per year, and the interferometer plane precesses 
around the normal to the Earth's orbit once per year.  Most sources will
be observed for a year or longer, thereby making full use of these
modulations.

\section{Binaries Made of Neutron Stars, Black Holes, and Other Exotic Objects:
Inspiral, Collision, and Coalescence in the HF Band}
\label{coalescing_binaries}

The best understood of all gravitational-wave sources are coalescing,
compact binaries composed of neutron stars (NS) and black holes (BH).
These NS/NS, NS/BH, and BH/BH binaries may well become the ``bread and
butter'' of the LIGO/VIRGO and LISA diets.  

The Hulse-Taylor \cite{hulse_taylor,taylor} binary pulsar, 
PSR 1913+16, is an example of a NS/NS binary
whose waves could be measured by LIGO/VIRGO, if we were to wait long
enough.  

At present PSR 1913+16 has an orbital frequency of about 1/(8 hours)
and emits its waves predominantly at twice this frequency, roughly
$10^{-4}$ Hz, which is in LISA's low-frequency band (cf.\ Fig.\
\ref{fig:lisa_noise}).  
As a result of their loss of orbital energy to gravitational waves,
the PSR 1913+16 NS's are gradually spiraling inward at a rate that agrees 
with general relativity's prediction to 
within the measurement accuracy (a fraction of a percent) 
\cite{taylor}---a remarkable but indirect confirmation that
gravitational waves do exist and are correctly described by general
relativity.  For this discovery and other aspects of PSR 1913+16,
Princeton's Russell Hulse and Joseph Taylor have been awarded the Nobel Prize.

If we wait roughly
$10^8$ years, this inspiral will bring the waves into the LIGO/ VIRGO
high-frequency band. As the NS's continue their inspiral,
the waves will then sweep upward in frequency, over a time of about 15
minutes, from 10 Hz to $\sim 10^3$ Hz, at which point the NS's will
collide and merge.  It is this last 15 minutes 
of inspiral, with $\sim 16,000$
cycles of waveform oscillation, and the final merger,
that LIGO/VIRGO seeks to monitor.

\subsection{Wave Strengths Compared to LIGO Sensitivities}
\label{cbsensitivities}

Figure \ref{fig:ligosources} above 
compares the projected sensitivities of interferometers in
LIGO \cite{ligoscience} 
with the wave strengths from the last few minutes of inspiral
of BH/BH, NS/BH, and NS/NS binaries at various distances from Earth.  
Notice that the signal strengths in Fig.\  \ref{fig:ligosources}
are in good accord with our rough estimates based on Eq.\ (\ref{hom});
at the endpoint (right end) of each inspiral, the number of cycles $n$ spent
near that frequency is of order unity, so the quantity plotted, $h_c
\simeq h\sqrt
n$, is about equal to $h$---and for a NS/NS binary at 200 Mpc is roughly 
$10^{-21}$,
as we estimated in Section \ref{armlength}.  

\subsection{Coalescence Rates}
\label{coalescencerates}

Such final coalescences are few and far between in our own galaxy: 
about one every 100,000 years, according to 1991 
estimates by Phinney \cite{phinney} and
by Narayan, Piran, and Shemi \cite{narayan}, 
based on the statistics of binary pulsar searches
in our galaxy which found three that will coalesce in less than
$10^{10}$ years.  Extrapolating out through the universe, 
Phinney and Narayan et.\ al.\ infer that to see several
NS/NS coalescences per year, LIGO/VIRGO will have to look out to a
distance of about 200 Mpc (give or take a factor $\sim 2$); cf.\  
Fig.\ \ref{fig:ligosources}.
Since these estimates were made, the binary pulsar searches have been
extended through a significantly larger volume of the galaxy than
before, and no new ones with coalescence times $\alt 10^{10}$ years have
been found.  This would drive the estimated event rate downward, except that
revisions of other aspects of the Phinney/Narayan analyses have compensated,
leaving the binary-pulsar-search-based
best estimates unchanged, at several events per year out to 200 Mpc 
\cite{vdheuvellorimer}.

A rate of one every 100,000 years 
in our galaxy is $\sim 100$ times smaller than the birth rate of 
the NS/NS binaries' progenitors: 
massive, compact, main-sequence binaries \cite{phinney,narayan}.
Therefore, either 99 per cent of progenitors
fail to make it to the NS/NS state (e.g., because of binary disruption
during a supernova or forming T\.ZO's), 
or else they do make it, but they wind up as a
class of NS/NS binaries that has not yet been discovered in any of the
pulsar searches.  Several experts on binary evolution have argued for
the latter \cite{tutukov_yungelson,yamaoka,lipunov}: most
NS/NS binaries, they suggest, may form with such short orbital periods
that their lifetimes to coalescence are significantly shorter than
normal pulsar lifetimes ($\sim 10^7$ years); and with such short
lifetimes, they have been missed in pulsar searches.  By modeling the
evolution of the galaxy's binary star population, these binary experts arrive at
best estimates as high as $3\times 10^{-4}$ coalescences per year in our
galaxy, corresponding to several per year out to 60 Mpc distance
\cite{tutukov_yungelson}, though more recent and more conservative models 
\cite{zwartspreeuw} give
results more nearly in accord with the binary pulsar searches, several per
year out to 200Mpc.  Phinney \cite{phinney} describes other
plausible populations of NS/NS binaries that could increase the event
rate, and he argues for ``ultraconservative'' lower and upper limits of 
23 Mpc and 1000Mpc
for how far one must look to see several coalescence per year. 

By comparing these rate estimates with the signal strengths in Fig.\  
\ref{fig:ligosources}, we see that: (i) The first interferometers
in LIGO/VIRGO (ca. 2001) have a possibility but not high probability of seeing
NS/NS coalescences. (ii) Enhanced interferometers (mid 2000's) can be fairly
confident of seeing them; the conservatively estimated event rate is $\sim 3
\times (300/200)^3 \simeq 10/$year.  (iii) Advanced interferometers are almost 
certain of seeing them (the requirement that
this be so was one factor that forced the LIGO/VIRGO arm lengths to be
so long, several kilometers). 
 
We have no good observational handle on the coalescence rate of NS/BH or
BH/BH binaries.  However, theory suggests that their progenitors might
not disrupt during the stellar collapses that produce the NS's and BH's,
so their coalescence rate could be about the same as the birth
rate for their progenitors: $\sim 1/100,000$ years in our galaxy.  This
suggests that within 200 Mpc distance there might be several NS/BH or
BH/BH coalescences per year.  
\cite{phinney,narayan,tutukov_yungelson,lipunov}.
This estimate should be regarded as a
plausible upper limit on the event rate and lower limit on the distance to
look \cite{phinney,narayan}.

If this estimate is correct, then NS/BH and BH/BH binaries will be seen
before NS/NS, and might be seen by the first LIGO/VIRGO interferometers
or soon thereafter \cite{hughes_flanagan}; cf.\ Fig.\   \ref{fig:ligosources}.
However, this estimate is far less certain than the
(rather uncertain) NS/NS estimates!

Once coalescence waves have been discovered, each further improvement of
sensitivity by a factor 2 will increase the event rate by $2^3 \simeq
10$.  Assuming a rate of several NS/NS per year at 200 Mpc, the advanced
interferometers of Fig.\  \ref{fig:ligosources} should see
$\sim 100$ per year.

\subsection{Inspiral Waveforms and the Information They Can Bring}
\label{cbwaveforms}

Neutron stars and black holes have such intense self gravity that it is
exceedingly difficult to deform them.  Correspondingly, as they spiral
inward in a compact binary, they do not gravitationally deform each other
significantly until several orbits before their final
coalescence \cite {kochanek,bildsten_cutler}.  This means
that the inspiral waveforms are determined to high accuracy by 
only a few, clean parameters:
the masses and spin angular momenta of the bodies, and the initial
orbital elements (i.e.\ the elements when the waves enter the LIGO/VIRGO band). 

\begin{figure}
\vskip 14.6pc
\special{hscale=65 vscale=65 hoffset=5 voffset=-13
psfile=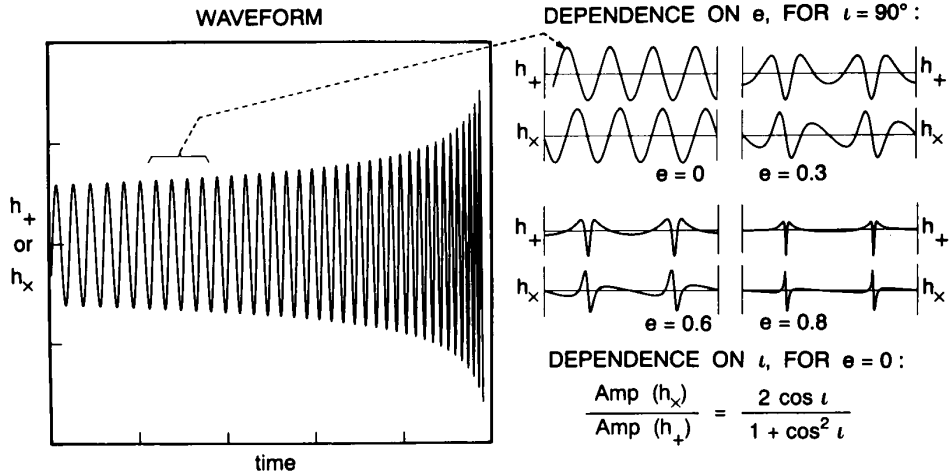}
\caption{Waveforms from the inspiral of a compact binary, computed using
Newtonian gravity for the orbital evolution and the quadrupole-moment
approximation for the wave generation.  (From Ref.\ 
\protect\cite{ligoscience}.)}
\label{fig:NewtonInspiral}
\end{figure}

Though tidal deformations are negligible during inspiral, relativistic
effects can be very important. 
If, for the moment, we ignore the relativistic effects---i.e., if we
approximate gravity as Newtonian and the wave generation as due to the
binary's oscillating quadrupole moment \cite{300yrs}, 
then the shapes of the inspiral
waveforms $h_+(t)$ and $h_\times(t)$
are as shown in Fig.\  \ref{fig:NewtonInspiral}.

The left-hand graph in Fig.\  \ref{fig:NewtonInspiral} shows the waveform 
increasing in
amplitude and sweeping upward in frequency 
(i.e., undergoing a ``chirp'') 
as the binary's bodies spiral closer and closer together.  The ratio of 
the amplitudes
of the two polarizations is determined by the inclination $\iota$ of the
orbit to our line of sight (lower right in Fig.\ \ref{fig:NewtonInspiral}).  The
shapes of the individual waves, i.e.\ the waves' harmonic content, are
determined by the orbital eccentricity (upper right).  (Binaries
produced by normal stellar evolution should be highly circular due to
past radiation reaction forces, but compact
binaries that form by capture events, in dense star clusters that might
reside in galactic nuclei \cite{quinlan_shapiro}, could be quite 
eccentric.)  If, for simplicity, the 
orbit is circular, then the rate at which 
the frequency sweeps or ``chirps'', $df/dt$ 
[or equivalently the number of cycles
spent near a given frequency, $n=f^2(df/dt)^{-1}$] is determined solely, in the
Newtonian/quadrupole approximation, by the binary's so-called {\it
chirp mass}, $M_c \equiv (M_1M_2)^{3/5}/(M_1+M_2)^{1/5}$ (where $M_1$ 
and $M_2$ are the two bodies' masses). 
The amplitudes of the two waveforms are determined by the chirp mass,
the distance to the source, and the orbital inclination.  Thus 
(in the Newtonian/quadrupole
approximation), by measuring the two amplitudes, the frequency sweep, and
the harmonic content of the inspiral waves, one can determine as direct,
resulting observables, the source's distance, chirp mass, inclination,
and eccentricity \cite{schutz_nature86,schutz_grg89}.

As in binary pulsar observations \cite{taylor}, 
so also here, relativistic effects add further information:
they influence the rate of frequency sweep and produce waveform
modulations in ways that
depend on the binary's dimensionless ratio $\eta = \mu/M$ of reduced mass 
$\mu = M_1 M_2/(M_1 + M_2)$ to total mass $M = M_1 + M_2$ 
and on the spins of the binary's two bodies. 
These relativistic effects are reviewed and discussed at length in 
Refs.\ \cite{last3minutes,will_nishinomiya}.  Two deserve special
mention: (i) As the waves emerge from the binary, some of them get
backscattered one or more times off the binary's spacetime curvature,
producing wave {\it tails}.  These tails act back on the binary,
modifying its inspiral rate in a measurable way.  (ii)  If the orbital 
plane is inclined to one or both of the binary's spins, then 
the spins drag inertial frames in the binary's
vicinity (the ``Lense-Thirring effect''), this frame dragging causes
the orbit to precess, and the precession modulates the 
waveforms \cite{last3minutes,precess,kidder}.  

Remarkably, the relativistic corrections to the frequency sweep --- tails, 
spin-induced precession and others --- will be 
measurable with rather high
accuracy, even though they are typically $\alt 10$ per cent of the 
Newtonian contribution, and even though the
typical signal to noise ratio will be only $\sim 9$. 
The reason is as 
follows \cite{cutler_flanagan,finn_chernoff,last3minutes}:

The frequency sweep will be monitored by the method of ``matched
filters''; in other words, the incoming, noisy signal will be cross 
correlated with theoretical templates.  If the signal and the templates
gradually 
get out of phase with each other by more than $\sim 1/10$ cycle as the 
waves sweep
through the LIGO/VIRGO band, their cross correlation will be significantly
reduced.
Since the total number of cycles spent in the LIGO/VIRGO band will
be $\sim 16,000$ for a NS/NS binary, $\sim 3500$ for NS/BH, and $\sim
600$ for BH/BH, this means that LIGO/VIRGO should be able to measure the
frequency sweep to a fractional precision $\alt 10^{-4}$, 
compared to which the relativistic effects
are very large.  (This is essentially the same method as
Joseph Taylor and colleagues use for high-accuracy radio-wave measurements of
relativistic effects in binary pulsars \cite{taylor}.)

Preliminary analyses, using the theory of optimal signal processing,
predict the following typical accuracies for LIGO/VIRGO measurements
based solely on the frequency sweep (i.e., ignoring modulational
information) 
\cite{poisson_will}: (i) The chirp mass $M_c$ 
will typically be measured, from the Newtonian part of the frequency
sweep, to $\sim 0.04\%$ for a NS/NS binary and 
$\sim 0.3\%$ for a system containing at least one BH. 
(ii) {\it If} we
are confident (e.g., on a statistical basis from measurements of many
previous binaries) that the spins are a few percent or less
of the maximum physically allowed, then the reduced mass $\mu$ 
will be measured to
$\sim 1\%$ for NS/NS and NS/BH binaries, and
$\sim 3\%$ for BH/BH binaries.  (Here and below NS means a
$\sim 1.4 M_\odot$
neutron star and BH means a $\sim 10 M_\odot$
black hole.) (iii) Because the
frequency dependences
of the (relativistic) $\mu$ effects 
and spin effects are not 
sufficiently different
to give a clean separation between $\mu$ and the spins,
if we have no prior knowledge of the spins, then 
the spin$/\mu$ correlation will
worsen the typical accuracy of $\mu$ by a large factor,
to $\sim 30\%$ for NS/NS, $\sim 50\%$ for NS/BH, and 
a factor $\sim 2$ for BH/BH. 
These worsened accuracies might be improved somewhat
by waveform modulations caused by the
spin-induced precession of the orbit \cite{precess,kidder}, 
and even without modulational information, a certain
combination of $\mu$ and the spins
will be determined to a few per cent.  Much
additional theoretical work is needed
to firm up the measurement accuracies.

To take full advantage of all the information in the inspiral waveforms
will require theoretical templates that are accurate, for given masses
and spins, to a fraction of a cycle during the entire sweep through the
LIGO/VIRGO band.  Such templates are being computed by an international
consortium of relativity theorists (Blanchet and Damour in France, Iyer
in India, Will and Wiseman in the U.S.)
\cite{2pnresults}, using post-Newtonian expansions of
the Einstein field equations.  This enterprise is rather like computing
the Lamb shift to high order in powers of the fine structure
constant, for comparison with experiment.  Cutler
and Flanagan \cite{cutler_flanagan1} have estimated the order to which the
computations must be carried in order that
systematic errors in the theoretical templates will not significantly
impact the information extracted from the LIGO/VIRGO observational data.
The answer appears daunting: radiation-reaction effects must be computed
to three full post-Newtonian orders [six orders in $v/c =$(orbital
velocity)/(speed of light)] beyond the leading-order radiation reaction,
which itself is 5 orders in $v/c$ beyond the Newtonian theory of
gravity---though by judicious use of Pad\'e approximates, these requirements
might be relaxed \cite{damour_sathya}.

It is only about ten years since controversies over the leading-order
radiation reaction \cite{quadrupole_controversy} were resolved by a 
combination of theoretical
techniques and binary pulsar observations.  Nobody dreamed then that
LIGO/VIRGO observations may require pushing post-Newtonian computations  
onward from $O[(v/c)^5]$ to $O[(v/c)^{11}]$.  This requirement epitomizes
a major change in the field of relativity research: At last, 80 years
after Einstein formulated general relativity, experiment has become a
major driver for theoretical analyses.

Remarkably, the goal of $O[(v/c)^{11}]$ is achievable.  The most difficult
part of the computation, the radiation reaction, has been evaluated to
$O[(v/c)^9]$ beyond Newton by the French/Indian/American consortium
\cite{2pnresults} 
and $O[(v/c)^{10}]$ is coming under control.

These high-accuracy waveforms are needed only for extracting information
from the inspiral waves, after the waves have been discovered; they are
not needed for the discovery itself.  The discovery is best achieved
using a different family of theoretical waveform templates, one that 
covers the space of potential waveforms
in a manner that minimizes computation time instead
of a manner that ties quantitatively into general relativity 
theory \cite{last3minutes,owen}.  Such templates are in the early stage of
development. 

\subsection{Testing GR, Measuring the Cosmological Universe, 
Mapping Black Holes, and
Searching for Exotic Objects}
\label{exotic_objects}

LIGO/VIRGO observations of compact binary inspiral have the potential to
bring us far more information than just binary masses and spins:
\begin{itemize}
\item 
They can be used for high-precision tests of general relativity.  In
scalar-tensor theories (some of which are attractive alternatives
to general relativity \cite{damour_nordvedt}), radiation reaction 
due to emission
of scalar waves places a unique signature on the gravitational
waves that LIGO/VIRGO
would detect---a signature that can be searched for with high precision
\cite{will_scalartensor}.
\item
They can be used to measure the Universe's Hubble constant, deceleration
parameter, and cosmological constant 
\cite{schutz_nature86,schutz_grg89,markovic,chernoff_finn}.  The keys to
such measurements are that: (i) Advanced interferometers in
LIGO/VIRGO will be able to see NS/NS
out to cosmological redshifts $z \sim 0.3$, and NS/BH out to $z
\sim 2$. (ii) The direct observables that can be extracted
from the
observed waves include the source's luminosity distance $r_{\rm L}$ (measured
to accuracy $\sim 10$ per cent in a large fraction of cases), and its
direction on the sky (to accuracy $\sim 1$ square degree)---accuracies
good enough that only one or a few electromagnetically-observed
clusters of galaxies should fall within the 3-dimensional
gravitational error boxes, thereby giving promise to joint
gravitational/electromagnetic statistical studies.  (iii) Another direct
gravitational observable is $(1+z)M$
where $z$ is redshift and $M$ is any mass in the system (measured to the
accuracies quoted above). Since the masses of NS's in binaries seem to
cluster around $1.4 M_\odot$, measurements of $(1+z)M$ can provide a
handle on the redshift, even in the absence of electromagnetic aid. 
\item
For a NS or small BH spiraling into a massive $\sim 50$ to $500 M_\odot$
compact central body, the inspiral waves will carry a ``map'' of the 
massive body's spacetime geometry.  This map can be used to determine whether 
the massive body is a black hole (in which case the geometry must be that of
Kerr, with all its features uniquely determined by its mass and angular
momentum) or is some other kind of exotic compact object, e.g. a soliton star
or naked singularity \cite{ryan,ryan_finn_thorne}. 
As we shall see in Section \ref{lfbhinspiral} below, this type of black-hole
study and search for exotic objects can be carried out with much higher 
precision
by LISA at large masses and low frequencies, than by LIGO/VIRGO at low masses
and high frequencies.
\end{itemize}

\subsection{Coalescence Waveforms and their Information}
\label{coalescence_waves}

The waves from the binary's final coalescence can bring us new
types of information.  

\bigskip
\centerline{\it BH/BH Coalescence}
\medskip

In the case of a BH/BH binary, the coalescence
will excite large-amplitude, highly nonlinear vibrations of spacetime
curvature near the coalescing black-hole horizons---a phenomenon of
which we have very little theoretical understanding today.  Especially
fascinating will be the case of two spinning black holes whose spins are
not aligned with each other or with the orbital angular momentum.  Each
of the three angular momentum vectors (two spins, one orbital) will drag
space in its vicinity into a tornado-like swirling motion---the general
relativistic ``dragging of inertial frames,'' so the binary is rather
like two tornados with orientations skewed to each other, embedded inside a 
third, larger tornado with a third orientation.  The dynamical evolution of 
such a complex configuration of coalescing spacetime warpage 
(as revealed by its
emitted waves) might bring us
surprising new insights into relativistic gravity \cite{ligoscience}.  
Moreover, if the sum
of the BH masses is fairly large, $\sim 40$ to $200 M_\odot$, then the waves
should come off in a frequency range $f\sim 40$ to $200$ Hz where the 
LIGO/VIRGO broad-band interferometers have their best sensitivity and can best 
extract
the information the waves carry.  

To get full value out of such wave observations will require 
\cite{hughes_flanagan} having
theoretical computations with which to compare them.
There is no hope to perform such computations
analytically; they can only be done as supercomputer simulations.
The development of such simulations
is being pursued by several research groups, including an 
eight-university American consortium of numerical relativists and 
computer scientists called the
Two-Black-Hole Grand Challenge Alliance \cite{GC} I have a bet with
Richard Matzner, the lead PI of this alliance, that LIGO/VIRGO will discover
waves from such coalescences with misaligned spins before the Alliance 
is able to compute them.

\bigskip
\centerline{\it NS/NS Coalescence}
\medskip

The final coalescence of NS/NS binaries should produce waves that are
sensitive to the equation of state of nuclear matter, so
such coalescences have the potential to teach us about the
nuclear equation of state \cite{ligoscience,last3minutes}.  In essence, 
LIGO/VIRGO will be 
studying nuclear physics via the collisions of atomic nuclei that have
nucleon numbers $A \sim 10^{57}$---somewhat larger than physicists are normally 
accustomed to. 
The accelerator used to drive these nuclei up to the speed of light is
the binary's self gravity, and the radiation by which the details of the
collisions are probed is gravitational.

Unfortunately, the final NS/NS coalescence will emit its gravitational
waves in the kHz frequency band ($800 {\rm Hz} \alt f \alt 2500 {\rm
Hz}$) where photon shot noise will prevent them from being studied by
the standard, ``workhorse,'' broad-band
interferometers of Fig.\  \ref{fig:ligosources}.
However, a specially configured (``dual-recycled'')
interferometer,
which could have enhanced sensitivity in the kHz
region at the price of reduced sensitivity elsewhere, may be able to
measure the waves and extract their equation of state information,
as might massive, spherical, resonant-mass detectors 
\cite{last3minutes,kennefick_laurence_thorne}. Such measurements will
be very difficult and are likely only when the LIGO/VIRGO
network has reached a mature stage.  

A number of research groups \cite{centrella}
are engaged in numerical astrophysics
simulations of NS/NS coalescence, with the goal not only to predict the
emitted gravitational waveforms and their dependence on equation of
state, but also (more immediately) to learn whether such 
coalescences
might power the $\gamma$-ray bursts that have been a major astronomical
puzzle since their discovery in the early 1970s.  

NS/NS coalescence is
a popular explanation for the $\gamma$-ray bursts because 
(i) the bursts are isotropically distributed on the sky, (ii) they have
a distribution of number versus intensity that suggests they might lie
at near-cosmological distances, and (iii) their event rate is
roughly the same as that predicted for NS/NS coalescence ($\sim1000$
per year out to cosmological distances, if they are cosmological).
If LIGO/VIRGO were now in operation and observing
NS/NS inspiral, it could
report definitively whether or not the $\gamma$-bursts are produced by
NS/NS binaries; and if the answer were yes, then the combination of
$\gamma$-burst data and gravitational-wave data could bring valuable
information that neither could bring by itself.  For example, it would
reveal when, to within a few msec, the $\gamma$-burst is emitted 
relative to the moment the NS's first begin to touch; and by
comparing the $\gamma$ and gravitational times of arrival, 
we might test whether gravitational waves propagate with
the speed of light to a fractional precision of 
$\sim 0.01{\rm sec}/3\times10^9\, {\rm lyr} = 10^{-19}$.

\bigskip
\centerline{\it NS/BH Coalescence}
\medskip

A NS spiraling into a BH of mass $M \agt 10
M_\odot$ should be swallowed more or less whole.  However, if the BH is
less massive than roughly $10 M_\odot$, and especially if it is rapidly
rotating, then the NS will tidally disrupt before being swallowed.
Little is known about the disruption and accompanying waveforms.  To
model them with any reliability will likely require full numerical
relativity, since the circumferences of the BH and NS will be comparable
and their physical separation at the moment of disruption
will be of order their separation. As with NS/NS, the coalescence 
waves should
carry equation of state information and will come out in the kHz band,
where their detection will require advanced, specialty detectors. 

\bigskip
\centerline{\it Christodoulou Memory} 
\medskip

As the coalescence waves depart from
their source, their energy creates (via the nonlinearity of Einstein's
field equations) a secondary wave called the ``Christodoulou memory''
\cite{christodoulou,thorne_memory,wiseman_will_memory}.  Whereas the primary
waves may have frequencies in the kHz band, the memory builds up on the
timescale of the primary energy emission profile, which is likely to be
of order 0.01 sec, corresponding to a memory frequency in the optimal
band for the LIGO/VIRGO workhorse interferometers, $\sim 100$Hz.  
Unfortunately, the memory is so weak that only very advanced
ground-based interferometers have much chance of detecting and studying 
it---and
then, perhaps only for BH/BH coalescences and not for NS/NS or NS/BH
\cite{kennefick_memory}.  LISA, by contrast, should easily be able to measure
the memory from supermassive BH/BH coalescences.

\section{Other High-Frequency Sources}
\label{otherhfsources}

\subsection{Stellar Core Collapse and Supernovae}
\label{supernovae}

When the core of a massive star has exhausted its supply of nuclear fuel, 
it collapses to form a neutron star or black hole. In some cases, the
collapse triggers and powers a subsequent explosion 
of the star's mantle---a supernova explosion.  Despite extensive
theoretical efforts for more than 30 years, and despite wonderful
observational data from Supernova 1987A, theorists are still far from  
a definitive understanding of the details of the collapse and explosion.  The
details are highly complex and may differ greatly from one
core collapse to another \cite{petschek}.  

Several features of the collapse and the core's subsequent
evolution can produce significant gravitational radiation in the 
high-frequency band. We shall
consider these features in turn, the most weakly radiating first. 

\bigskip
\centerline{\it Boiling of the Newborn Neutron Star}
\medskip

Even if the collapse is spherical, so it cannot radiate any
gravitational waves at all, it should 
produce a convectively unstable neutron
star that ``boils'' vigorously (and nonspherically) for the first 
$\sim 1$ second of its life \cite{bethe}.  The boiling dredges 
up high-temperature  
nuclear matter ($T\sim 10^{12}$K) from the neutron star's central regions,
bringing it to the surface (to the ``neutrino-sphere''), where it 
cools by
neutrino emission before being swept back downward and reheated.  Burrows
\cite{burrows1} has pointed out that the  boiling
should generate $n \sim 100$ cycles of gravitational waves with
frequency $f\sim 100$Hz and amplitude large enough to be detectable by
LIGO/VIRGO throughout our galaxy and its satellites.
Neutrino
detectors have a similar range, and there could be a high scientific payoff
from
correlated observations of the gravitational waves emitted by the
boiling's mass motions and neutrinos emitted from the boiling
neutrino-sphere.  With neutrinos to trigger on, the sensitivities of LIGO
detectors should be about twice as good as shown in Fig.\
\ref{fig:ligosources}. 

Recent 3+1 dimensional simulations by 
M\"uller and Janka \cite{muller_janka} suggest an rms
amplitude
$h \sim 2 \times 10^{-23} (20{\rm kpc}/r)$ (where $r$ is the distance to
the source), corresponding to a characteristic amplitude $h_c \simeq
h\sqrt n \sim 2\times 10^{-22} (20{\rm kpc}/r)$; cf.\ Fig.\ 
\ref{fig:ligosources}.  (The older 2+1 dimensional simulations give $h_c$ 
about 6 times larger than this \cite{muller_janka}, but presumably are less 
reliable.)  
LIGO should be 
able to detect such waves throughout our galaxy with an amplitude signal to 
noise ratio of about 
$S/N = 2.5$ in each of its two enhanced 4km interferometers, and its advanced
interferometers should do the same out to 80Mpc distance. 
(Recall that the $h_{\rm
SB}$ curves in Fig.\ \ref{fig:ligosources} are drawn at a signal to noise ratio
of about 5).  Although the event rate is only about one every
40 years in our galaxy and one every 20 years out to 80Mpc, the 
correlated neutrino and gravitational wave observations could bring very
interesting insights into the boiling of a newborn neutron star.

\bigskip
\centerline{\it Axisymmetric Collapse, Bounce, and Oscillations}
\medskip

Rotation will centrifugally flatten the collapsing
core, enabling it to radiate as it implodes.  If the core's angular
momentum is small enough that centrifugal forces do 
not halt or strongly slow the collapse before it reaches
nuclear densities, then the core's collapse, bounce, and subsequent
oscillations are likely to be axially symmetric.  Numerical
simulations \cite{finn_collapse,monchmeyer} show that in this case the waves
from collapse, bounce, and oscillation 
will be quite weak: the total energy radiated as gravitational waves
is not likely to exceed $\sim 10^{-7}$ solar masses (about 1 part in a
million of the collapse energy) and might often be much less than
this; and correspondingly, the waves' 
characteristic amplitude will be $h_c \alt 3\times 10^{-21}(30{\rm
kpc}/r)$.  These collapse-and-bounce waves will come off at frequencies
$\sim 200$ Hz to $\sim 1000$ Hz, and will precede the boiling waves by a
fraction of a second.  Though a little stronger than the boiling waves, 
they probably cannot be seen by LIGO/VIRGO beyond the local group of 
galaxies and thus will be a very rare occurrence.

\bigskip
\centerline{\it Rotation-Induced Bars and Break-Up}
\medskip

If the core's rotation is large enough to strongly flatten the 
core before or as it reaches nuclear density, 
then a dynamical and/or
secular instability is likely to break the core's axisymmetry.
The core will be transformed into
a bar-like configuration that spins end-over-end like
an American football, and that might even break up into two or more massive
pieces.  In this case, the radiation from the spinning bar or orbiting
pieces {\it could} be almost as strong as that from a coalescing neutron-star
binary, and thus could be seen by the LIGO/VIRGO first interferometers
out to the distance of the Virgo cluster (where the supernova rate is 
several per
year), by enhanced interferometers out to $\sim 100$Mpc (supernova rate several
thousand per year), and by advanced interferometers out to several hundred 
Mpc (supernova rate $\sim \hbox{(a few)}\times \sim 10^4$ per year); cf.\ 
Fig.\  \ref{fig:ligosources}.  It is far 
from clear what fraction of collapsing cores will have enough angular
momentum to break their axisymmetry, and what fraction of those will
actually radiate at this high rate; but even if only $\sim 1/1000$ or
$1/10^4$ do so, this could ultimately be a very interesting source for
LIGO/VIRGO.

Several specific scenarios for such non-axisymmetry have been identified:

{\bf Centrifugal hangup at $\bf \sim 100$km radius:} If the
pre-collapse core is rapidly spinning (e.g., if it is a white dwarf that
has been spun up by accretion from a companion), then the collapse may
produce a highly flattened, centrifugally supported disk with most of
its mass at radii $R\sim 100$km, which then (via instability)
may transform itself into a bar or may bifurcate.  The bar or
bifurcated lumps will radiate gravitational waves at twice their rotation 
frequency, $f\sim 100$Hz --- the optimal frequency for LIGO/VIRGO
interferometers.  To shrink on down to $\sim 10$km size, this
configuration must shed most of its angular momentum.  {\it If} a
substantial fraction of the angular momentum goes into
gravitational waves, then independently of the strength of the bar,
the waves will be nearly as strong as those from a coalescing binary.
The reason is this:
The waves' amplitude $h$ is proportional to the bar's ellipticity $e$,
the number of cycles $n$ of wave emission is proportional to $1/e^2$, and the
characteristic amplitude $h_c = h\sqrt n$ is thus independent of the
ellipticity and is about the same whether the configuration is a bar or
is two lumps \cite{schutz_grg89}.  The resulting waves will thus have $h_c$
roughly half as large, at $f\sim 100$Hz, as the $h_c$ from a NS/NS binary
(half as large because each lump might be half as massive as a NS), and
the waves will chirp upward in frequency in a manner similar to those from a
binary.

It is rather likely, however, that most of excess angular momentum does {\it
not} go into gravitational waves, but instead goes largely into hydrodynamic
waves as the bar or lumps, acting like a propeller, stir up the 
surrounding stellar mantle.  In this case, the radiation will be
correspondingly weaker. 

{\bf Centrifugal hangup at $\bf \sim 20$km radius:}  Lai and Shapiro
\cite{lai} have explored the case of centrifugal
hangup at radii not much larger than the final neutron star, say $R\sim
20$km.  Using compressible ellipsoidal models, they have deduced that,
after a brief period of dynamical bar-mode instability with wave
emission at $f\sim 1000$Hz (explored by
Houser, Centrella, and Smith \cite{houser}), the star switches to a secular
instability in which the bar's angular velocity gradually slows while
the material of which it is made retains its high rotation speed and
circulates through the slowing bar.  The slowing bar emits waves that sweep
{\it downward} in frequency through the LIGO/VIRGO optimal band $f\sim 100$Hz,
toward $\sim 10$Hz. The characteristic amplitude (Fig.\
\ref{fig:ligosources}) is only modestly smaller than for the upward-sweeping
waves from hangup at $R\sim 100$km, and thus such waves should be
detectable near the Virgo Cluster by the first LIGO/VIRGO interferometers,
near 100Mpc by enhanced interferometers, and 
and at distances of a few 100Mpc by advanced interferometers.

{\bf Successive fragmentations of an accreting, newborn neutron star:}
Bonnell and Pringle \cite{pringle} have focused on the evolution of the
rapidly spinning, newborn neutron star as it quickly accretes more and
more mass from the pre-supernova star's inner mantle.  If the accreting
material carries high angular momentum, it may trigger a renewed bar
formation, lump formation, wave emission, and coalescence, followed by more
accretion, bar and lump formation, wave emission, and coalescence.  Bonnell
and Pringle
speculate that hydrodynamics, not wave emission, will drive this
evolution, but that the total energy going into gravitational waves might be
as large as $\sim 10^{-3}M_\odot$.  This corresponds to $h_c \sim 10^{-21}
(10{\rm Mpc}/r)$.

\subsection{Spinning Neutron Stars; Pulsars}
\label{pulsars}

As the neutron star settles down into its final state, its crust begins
to solidify (crystalize). The solid
crust will assume nearly the oblate axisymmetric shape that 
centrifugal forces are trying to maintain,
with poloidal 
ellipticity $\epsilon_p \propto$(angular velocity of rotation)$^2$. 
However, the principal axis
of the star's moment of inertia tensor may deviate from its spin axis
by some small ``wobble angle'' $\theta_w$, and the star may 
deviate slightly from axisymmetry about its principal axis; i.e., it may
have a slight ellipticity $\epsilon_e \ll \epsilon_p$ in its equatorial plane.

As this slightly imperfect crust spins, it will radiate gravitational
waves \cite{zimmermann}: $\epsilon_e$ radiates at twice the rotation 
frequency, $f=2f_{\rm
rot}$ with
$h\propto \epsilon_e$, and the wobble angle couples to $\epsilon_p$ to
produce waves at $f=f_{\rm rot} + f_{\rm prec}$
(the precessional sideband of the rotation frequency) with amplitude
$h\propto \theta_w \epsilon_p$.  For typical neutron-star masses and
moments of inertia, the wave amplitudes are
\begin{equation}
h \sim 6\times 10^{-25} \left({f_{\rm rot}\over 500{\rm Hz}}\right)^2
\left({1{\rm kpc}\over r}\right)\left({\epsilon_e \hbox{ or }\theta_w\epsilon_p
\over 10^{-6}}\right)\;.
\label{hpulsar}
\end{equation}
  
The neutron star gradually spins down, due in part to gravitational-wave
emission but perhaps more strongly due to electromagnetic torques associated
with its spinning magnetic field and pulsar emission. 
This spin-down reduces the strength of centrifugal forces, and thereby
causes the star's poloidal ellipticity $\epsilon_p$ to decrease, with
an accompanying breakage and resolidification of its crust's crystal structure
(a ``starquake'') \cite{starquake}.  
In each starquake, $\theta_w$, $\epsilon_e$, and
$\epsilon_p$ will all change suddenly, thereby changing the amplitudes and
frequencies of the
star's two gravitational ``spectral lines'' $f=2f_{\rm rot}$ and
$f=f_{\rm rot} + f_{\rm prec}$.  After each quake, there should be a
healing period in which the star's fluid core and solid crust, now rotating
at different speeds, gradually regain synchronism.
By monitoring the 
amplitudes, frequencies, and phases of the two gravitational-wave
spectral lines, and by 
comparing with timing of
the electromagnetic pulsar emission, one might learn much about the 
physics of the neutron-star interior.

How large will the quantities $\epsilon_e$ and $\theta_w \epsilon_p$ be?
Rough estimates of the crustal shear moduli and breaking strengths suggest an
upper limit in the range $\epsilon_{\rm max} \sim 10^{-4}$
to $10^{-6}$, and it might be that typical values are 
far below this.  We are extremely ignorant, and
correspondingly there is much to be learned from searches for
gravitational waves from spinning neutron stars.

One can estimate the sensitivity of LIGO/VIRGO (or any other broad-band
detector)
to the periodic waves from such a source by multiplying the waves'
amplitude $h$ by the square root of the number of cycles over which one
might integrate to find the signal, $n= f \hat \tau$ where $\hat\tau$ is the
integration time.  The resulting
effective signal strength, $h\sqrt{n}$, is larger than $h$ by
\begin{equation}
\sqrt n = \sqrt{f\hat\tau} = 10^5 \left( {f\over1000{\rm Hz}}\right)^{1/2}
\left({\hat\tau\over4{\rm months}}\right)^{1/2}\;.
\label{ftau}
\end{equation}
Four months of integration is not unreasonable in targeted searches; but for an
all-sky, all-frequency search, a coherent integration might not last longer
than a few days because of computational limitations associated with
having to apply huge numbers of trial neutron-star spindown corrections and
earth-motion doppler corrections \cite{brady_creighton_cutler_schutz}.

Equation (\ref{ftau}) for $h\sqrt n$  should be compared (i) to the
detector's rms broad-band noise level for sources in a random direction,
$\sqrt5 h_{\rm rms}$, to deduce a
signal-to-noise ratio, or (ii) to $h_{\rm SB}$ to deduce a
sensitivity for
high-confidence detection when one does not know the waves' frequency in
advance \cite{300yrs}.   
Such a comparison suggests that the first interferometers in
LIGO/VIRGO might possibly see waves from nearby spinning
neutron stars, but the odds of success are very unclear.

The deepest searches for these nearly periodic waves will be
performed by narrow-band detectors, whose sensitivities are enhanced
near some chosen frequency at the price of sensitivity loss
elsewhere---e.g., dual-recycled interferometers \cite{meers} or resonant-mass
antennas (Section \ref{narrowband}).
With ``advanced-detector technology'' and targeted searches, dual-recycled 
interferometers
might be able to detect with confidence spinning neutron stars 
that have \cite{300yrs}
\begin{equation}
(\epsilon_e \hbox{ or } \theta_w\epsilon_p ) \agt 3\times10^{-10} \left(
{500 {\rm Hz}\over f_{\rm rot}}\right)^2 \left({r\over 1000{\rm pc}}\right)^2.
\label{advancedpulsar}
\end{equation}
There may well be a large number of such neutron stars in our galaxy; but
it is also conceivable that there are none.  We are extremely
ignorant.

Some cause for optimism arises from several physical mechanisms that
might generate radiating ellipticities large compared to
$3\times10^{-10}$: 
\begin{itemize}

\item It may be that, inside the superconducting cores of
many neutron stars, there are trapped magnetic fields with mean
strength $B_{\rm core}\sim10^{13}$G or even
$10^{\rm 15}$G. 
Because such a field is actually concentrated in flux
tubes with $B = B_{\rm crit} \sim 6\times 10^{14}$G surrounded by
field-free superconductor, its mean pressure is $p_B = B_{\rm core} B_{\rm
crit}/8\pi$.  This pressure could produce a radiating 
ellipticity  
$\epsilon_{\rm e} \sim \theta_w\epsilon_p \sim p_B/p \sim 10^{-8}B_{\rm
core}/10^{13}$G (where $p$ is the core's material pressure). 

\item Accretion onto a spinning neutron star can drive precession (keeping
$\theta_w$ substantially nonzero), and thereby might produce measurably strong
waves \cite{schutz95}.

\item If a neutron star is born rotating very rapidly,
then it may experience a
gravitational-radiation-reaction-driven instability.  In this
``CFS'' (Chandrasekhar, \cite{cfs_chandra} Friedman, Schutz
\cite{cfs_friedman_schutz}) instability,
density waves travel around the
star in the opposite direction to its rotation, but are dragged forward
by the rotation.  These density waves produce gravitational waves that 
carry positive energy as seen by observers far from the star, but
negative energy from the star's viewpoint; and because the
star thinks it is losing negative energy, its density waves get
amplified.  This intriguing mechanism is similar to that by which
spiral density waves are produced in galaxies.  Although the CFS
instability was once thought ubiquitous for spinning stars 
\cite{cfs_friedman_schutz,wagoner}, we now
know that neutron-star viscosity will kill it, stabilizing the star and
turning off the waves, when the star's temperature is above some
limit $\sim 10^{10}{\rm K}$ \cite{cfs_lindblom}
and below some limit $\sim 10^9 {\rm K}$
\cite{cfs_mendell_lindblom}; and correspondingly, the instability
should operate only during the first few years of a neutron
star's life, when $10^9 {\rm K} \alt T \alt 10^{10}\rm K$.

\end{itemize}

\section{Low-Frequency Gravitational-Wave Sources}
\label{lfsources}

\subsection{Waves from the Coalescence of Massive Black Holes in 
Distant Galaxies}
\label{lfbhcoalescence}

LISA would be a powerful instrument for studying massive black holes in
distant galaxies.  Figure \ref{fig:lisa_noise} shows, as examples, the
waves from several massive black hole binaries at 3Gpc distance from Earth (a
cosmological redshift of unity).  The waves sweep upward in frequency
(rightward in the diagram) as the holes spiral together.  The black dots
show the waves' frequency one year before the holes' final collision and
coalescence, and the arrowed lines show the sweep of frequency and
characteristic amplitude $h_c = h\sqrt n$ during that last year.  For
simplicity, the figure is restricted to binaries with equal-mass black
holes: 
$10^4M_\odot / 10^4 M_\odot$, $10^5M_\odot / 10^5 M_\odot$, 
and $10^6M_\odot / 10^6 M_\odot$.  

By extrapolation from these three examples, we see that LISA can
study much of the last year of inspiral, and the waves 
from the final collision and coalescence, 
whenever the holes' masses are in the range $3\times 10^4 M_\odot 
\alt M \alt 3\times 10^8 M_\odot$ \cite{hughes_flanagan}.  Moreover, 
LISA can study the final coalescences with remarkable 
signal to noise ratios: $S/N \agt 1000$.    
Since these are much larger $S/N$'s than LIGO/VIRGO is likely to achieve, 
we can expect LISA to refine the experimental understanding of black-hole
physics, and of highly nonlinear vibrations of warped spacetime, 
which LIGO/VIRGO initiates---{\it provided} the rate of massive
black-hole coalescences is of order one per
year in the Universe or higher.  The rate might well be that high, but 
it also might be much lower. 

By extrapolating Fig.\  \ref{fig:lisa_noise} to lower BH/BH masses, we
see that LISA can observe the last few years of inspiral, but not the
final collisions, of binary black holes in the range 
$100M_\odot \alt M \alt 10^4 M_\odot$, out to cosmological distances
\cite{hughes_flanagan}.

Extrapolating the BH/BH curves to lower frequencies using the
formula (time to final coalescence$)\propto f^{-8/3}$, we see that 
equal-mass BH/BH binaries enter LISA's frequency band roughly 1000 years
before their final coalescences, more or less independently of their
masses, for the range $100 M_\odot \alt M \alt 10^6 M_\odot$.  Thus, if the
coalescence rate were to turn out to be one per year, LISA would see
roughly 1000 additional massive binaries that are slowly spiraling
inward, with inspiral rates $df/dt$ readily measurable.  From the inspiral
rates, the amplitudes of the two polarizations, and the waves' harmonic
content, LISA can determine each such binary's luminosity distance,
redshifted chirp mass $(1+z)M_c$, orbital inclination, 
and eccentricity; and from the waves' modulation by LISA's orbital
motion, LISA can learn the direction to the binary with an accuracy of
order one degree. 

\subsection{Waves from Compact Bodies Spiraling into Massive Black
Holes or Exotic Objects in Distant Galaxies}
\label{lfbhinspiral}

When a compact body with mass $\mu$ spirals into a much more massive black
hole with mass $M$, the body's orbital energy $E$ at fixed frequency
$f$ (and correspondingly at fixed orbital radius $a$)
scales as $E \propto \mu$, 
the gravitational-wave luminosity $\dot E$ scales as 
$\dot E \propto \mu^2$, and the time to
final coalescence thus scales as $t \sim E/\dot E \propto 1/\mu$.  This   
means that the smaller is $\mu/M$,
the more orbits are spent in the hole's strong-gravity region, $a\alt
10GM/c^2$, and thus the more detailed and accurate will be the map of the
hole's spacetime geometry, which is encoded in the emitted waves.

For holes observed by LIGO/VIRGO, the most extreme mass ratio that we
can hope for is $\mu/M \sim 1M_\odot/300 M_\odot$, since for $M>300M_\odot$ the
inspiral waves are pushed to frequencies below the LIGO/VIRGO band.  
This limit on $\mu/M$ seriously constrains the accuracy with which
LIGO/VIRGO can hope to map out the spacetime geometries of black
holes and test the black-hole no-hair theorem \cite{ryan,ryan_finn_thorne} 
(end of Section \ref{exotic_objects}). 
By contrast, LISA can observe the final inspiral waves from objects of
any mass $M\agt 0.5M_\odot$ spiraling into holes of mass $3\times 10^5 M_\odot
\alt M \alt 3\times10^7M_\odot$.  

Figure \ref{fig:lisa_noise} shows the
example of a $10M_\odot$ black hole spiraling into a $10^6M_\odot$ hole
at 3Gpc distance.  The inspiral orbit and waves are strongly influenced
by the hole's spin.  Two cases are shown \cite{finn_thorne}: 
an inspiraling circular orbit 
around a non-spinning hole, and a prograde, circular, equatorial orbit 
around a maximally spinning hole. 
In each case the dot at the upper left end of the
arrowed curve is the frequency and characteristic amplitude one year
before the final coalescence.  In the nonspinning case, the small hole
spends its last year spiraling inward from $r\simeq 7.4 GM/c^2$ 
(3.7 Schwarzschild
radii) to its last stable circular orbit at $r=6GM/c^2$ (3 Schwarzschild
radii).  In the maximal spin case, the last year is spent traveling from
$r=6GM/c^2$ (3 Schwarzschild radii) to the last stable orbit at $r=GM/c^2$ 
(half a
Schwarzschild radius).  The $\sim 10^5$ cycles of waves during this last
year should carry, encoded in themselves, rather accurate values for 
the massive hole's lowest few multipole moments \cite{ryan} (or, equivalently,
a fairly accurate map of the hole's spacetime geometry (or,hole's).  If the
measured moments satisfy the ``no-hair'' theorem (i.e., if they are all
determined uniquely by the measured mass and spin in the manner of the
Kerr metric), then we can be sure the central body is a black hole.  If
they violate the no-hair theorem, then (assuming general relativity is
correct), either the central body was an exotic object (e.g. soliton star or
naked singularity) rather than a black hole, or else an accretion
disk or other material was perturbing its orbit \cite{chakrabarti}. 
From the evolution of the waves one can hope to determine which is
the case, and to explore the properties of the central body and its
environment \cite{ryan_finn_thorne}.

Models of galactic nuclei, where massive holes reside, suggest that
inspiraling stars and small holes typically will be in rather eccentric
orbits \cite{hils_bender}.  This is because they get injected into such 
orbits via
gravitational deflections off other stars, and by the time gravitational
radiation reaction becomes the dominant orbital driving force, there is
not enough inspiral left to fully circularize their orbits.  Such orbital
eccentricity will complicate the waveforms and complicate the extraction
of information from them.  Efforts to understand the emitted waveforms
are just now getting underway.  

The event rates for inspiral into massive black holes are not at all
well understood.  However, since a significant fraction of all galactic
nuclei are thought to contain massive holes, and since white dwarfs and
neutron stars, as well as small black holes, can withstand tidal 
disruption as they plunge toward the massive hole's horizon, and since
LISA can see inspiraling bodies as small as $\sim 0.5 M_\odot$ out to 
3Gpc distance, the event rate is likely to be interestingly large. 

\section{Conclusion}
\label{conclusion}

It is now 36 years since Joseph Weber initiated his pioneering
development of gravitational-wave detectors \cite{weber} and 25 years
since Forward \cite{forward} and Weiss \cite{weiss}
initiated work on interferometric detectors.  
Since then, hundreds of
talented experimental physicists have struggled to improve the
sensitivities of these instruments.  At last, success is in sight.  If
the source estimates described in this lecture are approximately
correct, then the planned interferometers should detect the first waves
in 2001 or several years thereafter, thereby opening up this rich new
window onto the Universe.  

\section{Acknowledgments}
\label{acknowledgments}

My group's research on gravitational waves 
and their relevance to LIGO/VIRGO and LISA is supported in part
by NSF grants AST-9417371 and PHY-9424337 and by NASA grant NAGW-4268.
This article was largely adapted and updated from my Ref.\ 
\cite{snowmass} and is a slightly updated version of my Ref.\ \cite{princeton}.


%
\end{document}